\documentclass[twocolumn,superscriptaddress,prl]{revtex4}
\usepackage{amsmath,amssymb}
\usepackage{graphicx}
\usepackage{color}

\usepackage{times}
\usepackage[colorlinks,bookmarks=false,citecolor=blue,linkcolor=red,urlcolor=blue]{hyperref}
\def \hf{\tfrac{1}{2}}

\def \ord{\mathcal{O}}

\begin{document}

\author{Vincenzo Alba}

\affiliation{Department of Physics and Arnold Sommerfeld Center for 
Theoretical Physics, Ludwig-Maximilians-Universit\"at M\"unchen, D-80333 M\"unchen, Germany}

\author{Masudul Haque}

\affiliation{Max-Planck-Institut f\"{u}r Physik komplexer Systeme, N\"{o}thnitzer Stra{\ss}e 38, D-01187 Dresden, Germany}

\author{Andreas~M.~L\"auchli}

\affiliation{Institut f\"ur Theoretische Physik, Universit\"at Innsbruck, A-6020 Innsbruck, Austria}

\date{\today}

\title{Entanglement Spectrum of the Two-Dimensional Bose-Hubbard Model}

\begin{abstract} 

We study the entanglement spectrum (ES) of the Bose-Hubbard model on the two dimensional square lattice at unit filling,
both in the Mott insulating and in the superfluid phase. In the Mott phase, we demonstrate that the ES is dominated by
the physics at the boundary between the two subsystems.  On top of the boundary-local (perturbative) structure, the ES
exhibits substructures arising from one-dimensional dispersions along the boundary.  In the superfluid phase, the
structure of the ES is qualitatively different, and reflects the spontaneously broken $U(1)$ symmetry of the phase.  We
attribute the basic low-lying structure to the ``tower of states'' (TOS) Hamiltonian of the model.  We then discuss how
these characteristic structures evolve across the superfluid to Mott insulator transition and their influence on the
behavior of the entanglement entropies.  We briefly outline the implications of the ES structure on the
efficiency of matrix-product-state based algorithms in two dimensions.

\end{abstract}

% \pacs{73.43.Cd, 71.10.Pm  {\tt check!}}

\maketitle

%\section{outline}
%\line(1,0){240}

%\line(1,0){240}

\paragraph*{Introduction ---}

In recent years the cross fertilization between quantum information and condensed matter has led to new insights into
the physics of low dimensional systems~\cite{AmicoFazioOsterlohVedral_RMP08}.  In particular the concept of
\emph{entanglement spectrum} (ES) has established itself as an informative and
intriguing theme.
Considering a bipartion of the system into parts $A$ and $B$, the ES, $\{\xi_i\}$, is defined in terms of the Schmidt
decomposition
\begin{equation}  \label{eq:schmidt_ES_defn}
|\psi\rangle=\sum_i e^{-\xi_i/2}|\psi_i^A\rangle
\otimes |\psi_i^B\rangle. 
\end{equation}
Here $|\psi\rangle$ is the ground state, and the states $|\psi_i^A\rangle$ ($|\psi_i^B\rangle$) form an orthonormal
basis for subsystem $A$ ($B$).  The ES $\{\xi_i=-\log\lambda_i\}$ is also the spectrum of the so called entanglement
hamiltonian ${\mathcal H}_E\equiv -\log\rho_A$ where the reduced density matrix $\rho_A$ is obtained after tracing out
the $B$ part of the system density matrix $|\psi\rangle\langle\psi|$.
The ES can also be used to construct \emph{entanglement entropies} (Renyi and von Neumann) which quantify the
entanglement between the two subsystems.

Many results are now available for the ES and entanglement entropies of one-dimensional (1D) systems.  In contrast,
higher dimensions are far less explored.  Much of the ES literature on 2D systems focuses on topological
phases \cite{topological}.  A detailed understanding of generic ES features for more common 2D systems is not currently
available.

In this work we present a thorough investigation of the ES for the 2D Bose-Hubbard model~\cite{fisher-89, jaksch-98}.
We extract features for both the gapped Mott insulator and the superfluid (gapless, symmetry-broken) phases.  This
analysis provides insights into the ES which, in addition to the intrinsic importance of the Bose-Hubbard model, should
be generic to gapped and symmetry-broken gapless 2D phases.  Our analysis is based on DMRG calculations on a cylindrical
geometry, complemented by perturbative calculations (Mott phase) and by analysis of the tower of states analogy
(superfluid phase).

%%%%%%%%%%%%%%%%%%%%%%%%%%%%%%%%%e
\begin{figure*}[t]
\includegraphics*[width=0.95\linewidth]{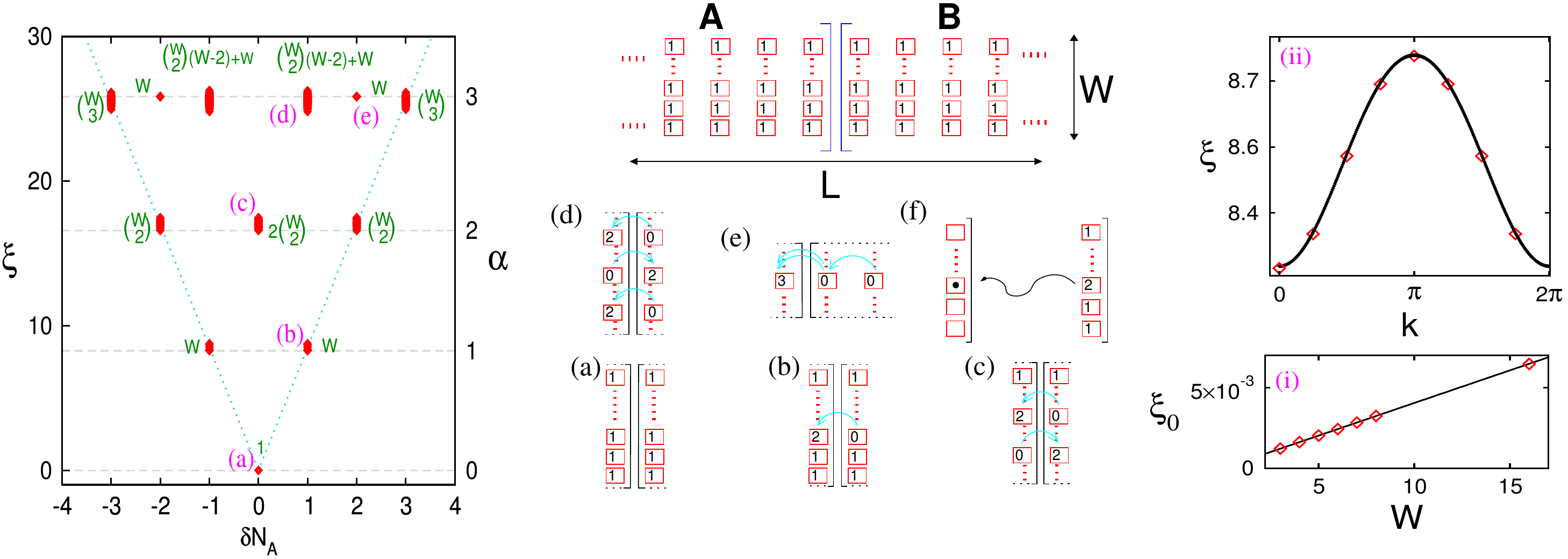}
\caption{  \label{fig1}
ES in the Mott phase.  ({\bf left}) Perturbative structure, DMRG data for equally bipartitioned cylinder, $W=8, L=32$,
$U=100$.
Multiplicities are indicated near each group of states The dotted lines highlight the ES envelope. ({\bf center bottom})
Bosonic configurations corresponding to leading order configurations for some $\delta{N_A}\ge0$ levels.  Only occupancies on sites
nearer the boundary are shown (omitted sites are singly occupied). 
The arrows represent hopping (perturbative) processes to create a given configuration. 
In $(d)$, to save space we show only one type of contributing configuration.  The shown and omitted
configurations contribute respectively ${\hf}W(W-1)(W-2)$ and $W$ to the multiplicity.
In $(f)$ we show the effective single particle (boundary) excitation corresponding to multiplet
$(b)$. ({\bf right}) Entanglement dispersions.  Data points are DMRG ($U=100$); black lines are
perturbative results.  (\textbf{i}) Lowest ES level $\xi_0$ as a function of $W$.  (\textbf{ii})
Entanglement dispersion for the lowest ES multiplet at $\delta N_A=1$.  $k$ denotes the momentum
along the boundary.
}
\end{figure*}
%%%%%%%%%%%%%%%%%%%%%%%%%%%%%%%%%%

In the 2D Mott insulator, we show that the ES is ``boundary-local'' in the sense introduced in Ref.~\cite{alba-2011} for
1D gapped systems: the ES is organized in a perturbative structure with higher ES levels corresponding to excitations
farther from the boundary.
In 2D the boundary is an extended object and not a single point, so that excitations created near the boundary can move
without changing the distance from the boundary.  In the ES, this shows up as `dispersive' structures within each order of
the perturbative hierarchy.
We present perturbative results for the corresponding entanglement Hamiltonians (${\mathcal H}_E$) and for entanglement
entropies.  We discuss how area laws emerge in the perturbative framework.

At the transition between the Mott and superfluid phases the ES structure changes dramatically.  We show that its lower
part (which we call ES ``envelope'') sharply reflects the different nature of the two phases.

An intriguing feature of the ES in phases with a broken continuous symmetry is the relation \cite{met-grov-2011} between
its lower part and the so-called tower of states (TOS) {\em energy} spectrum obtained when a system with continuous
symmetry breaking in the thermodynamic limit is placed in a finite volume \cite{lhuillier_arxiv2005}.  We demonstrate
that in the superfluid phase the ES envelope, separated from the rest of the spectrum by an apparent ``entanglement
gap'', reflects the appropriate TOS Hamiltonian.  This correspondence provides a framework to understand generic
features of the ES envelope.  This is an explicit demonstration of the TOS structure in the ES for a specific many-body
system in a symmetry-broken phase.  This paves the way toward the use of TOS analysis of symmetry-breaking based on
\emph{ground-state} calculations alone.  Conventional TOS analysis requires the calculation of properties of a
significant number of excited states of the full system \cite{lhuillier_arxiv2005}.

\paragraph*{The Bose-Hubbard model ---}

We consider the two dimensional Bose-Hubbard model on a cylinder of length $L$ (with open boundary conditions) and
circumference $W$ (with periodic boundary conditions).  The Hamiltonian is ${\mathcal H}=-\sum_{\langle ij\rangle}
(b^\dagger_ib_j+h.c.)+\frac{U}{2}\sum _{i}n_i(n_i-1)$ where $b_i$ are bosonic operators, $n_i=b^\dagger_ib_i$, and $U$
is the on-site repulsion. We restrict to the case of unit filling.  To calculate the ES we divide the cylinder in two
parts $A$ and $B$ of size $V_A$ and $V_B$ respectively, with $W$ the boundary length (Fig.~\ref{fig1} center top).  The
number of bosons $N_A(N_B)$ in $A(B)$ is a good quantum number for the ES and can be used to label the ES levels.  We
introduce $\delta N_A\equiv N_A- V_A$ measuring the excess of bosons in $A$ compared to unit filling.  
%

%%%%%%%%%%%%%%%%%%%%%%%%%%%%%%%%%%%%%%%%%%%%%

\paragraph*{The ES in the Mott insulator ---}

In the $U\to\infty$ limit, the ground state is a product state with one boson per site.  The ES has only one level,
$\xi_0=0$.  As in Ref.~\cite{alba-2011}, the ES at large $U$ can be constructed through boundary-linked perturbation
theory, treating the hopping term ${\mathcal H}_p\equiv-\sum_{\langle ij\rangle}(b^\dagger_ib_j+h.c.)$ as perturbation.

Fig.~\ref{fig1} (left) plots the ES $\{\xi\}$ for a system at $U=100$.  We denote with $\alpha$ the perturbative order
giving the leading contribution to the ES levels.  As in gapped 1D systems \cite{alba-2011}, the separation ($\sim\log
U$) between consecutive $\alpha$ reflects the perturbative nature of the ES.  In the center bottom we show the
configurations giving dominant contribution to selected groups of ES levels.  
%
%% The $A$ part of these configurations also gives the dominant configuration in the entanglement eigenfunctions
%% (eigenfunctions of $\rho_{A}$).
%
At each $\delta
N_A$ and $\alpha>0$ there is more than one level.  The multiplicities are determined by the boundary length $W$, and can
be understood in terms of hopping processes across the boundary.  On the ES envelope (dotted line in Fig.~\ref{fig1}
left), the group of levels at $\delta N_A$ is obtained by transferring $\delta N_A$ bosons from subsystem $B$ to $A$.
This process appears at order $\delta N_A$ (with amplitude $U^{-\delta N_A}$).  The multiplicity $m(\delta N_A)$ is
obtained as the number of ways of moving $\delta N_A$ bosons across the boundary to give distinct configurations, i.e.
$m(\delta N_A)=C(W,\delta N_A)$ with $C$ the binomial coefficient (see ({\bf a})({\bf b}) in Fig.~\ref{fig1} center).

Similar boundary perturbative processes help explain ES levels above the envelope; c.f., ({\bf c})({\bf d})({\bf e}) in
Fig.~\ref{fig1} left and center.  At $\alpha<3$ the leading order of all ES levels (not only the envelope) is given by
pure boundary processes. At higher orders ($\alpha{\geq}3$)  processes further away from the boundary start 
contributing to the leading order of some ES levels, e.g., $({\bf e})$ in Fig.~\ref{fig1}.  The corresponding configurations 
involve hoppings perpendicular to the boundary, and thus are similar to the physics of ES configurations in 1D 
chains \cite{alba-2011}.  This is reflected in the linear multiplicity (${\sim}W$) in ({\bf e}).  Generically, the multiplicities 
grow exponentially with $\alpha$ (${\sim}W^\alpha$).  This growth of multiplicity is related to the area law of entanglement 
entropy which limits the performance of DMRG in 2D systems even in gapped phases.

\paragraph{Entanglement dispersions \& entanglement Hamiltonian ---}
 
We now examine the substructures superposed on the perturbative hierarchy.  We first consider the ES states marked ({\bf
  b}) in Fig.~\ref{fig1}.  These $W$ states involve particle-hole excitations across the boundary at the $W$
possible positions along the boundary.  This local excess/depletion of bosons can be regarded as a boundary degree of
freedom [Fig~\ref{fig1}({\bf f})].  Since the momentum $k$ along the boundary is a good quantum number, in the
entanglement eigenstates these excitations will not be localized, but will appear as momentum eigenstates, similar to a
single-particle dispersion.  This ``entanglement dispersion'' is given perturbatively up to $\mathcal{O}(U^{-3})$ as
\begin{multline}
\label{tight}
\xi_{k}=\log(U^2/2)+2(2W+10)/U^2-24\Big[1/U+(2W-\\
25)/U^3\Big]\cos k-68/U^2\cos 2k+552/U^3\cos 3k.
\end{multline}
Comparison with DMRG data ($U=100$, $W=8$) is shown in Fig.~\ref{fig1}
right panel ({\bf ii}). 

The spectrum~\eqref{tight} can be interpreted as the spectrum of the ``Hamiltonian'' ${\mathcal H}_E=
\mathrm{const.}+\sum_{ir} A_r(a^\dagger_i a_{i+r}+h.c.)$ where $i$ labels sites of the boundary and $a_{i}$ are the
boundary degrees of freedom.  This is a boundary Hamiltonian of single particle tight-binding form, with hopping
amplitudes $A_r$ decaying exponentially with distance $r$ \cite{supplement, poil-2011}.  This can be generalized to each
group of states on the ES envelope: the envelope states in sector $\delta{N_A}$ are described by a boundary entanglement
Hamiltonian ${\mathcal H}_E$ for $\delta{N_A}$ particles.  Expressions for ${\mathcal H}_E$ can be obtained
perturbatively \cite{supplement}.  Up to lowest subleading order, ${\mathcal H}_E$ is of nearest neighbor tight-binding
form (on a chain).
As the Mott-superfluid transition is approached, ${\mathcal H}_E$ becomes more and more long range \cite{poil-2011,
supplement}, similar to the correlation length which diverges upon approaching the phase transition.

%%%%%%%%%%%%%%%%%%%%%%%%%%%%%%%%%%
\begin{figure}[t]
\includegraphics[width=1\columnwidth]{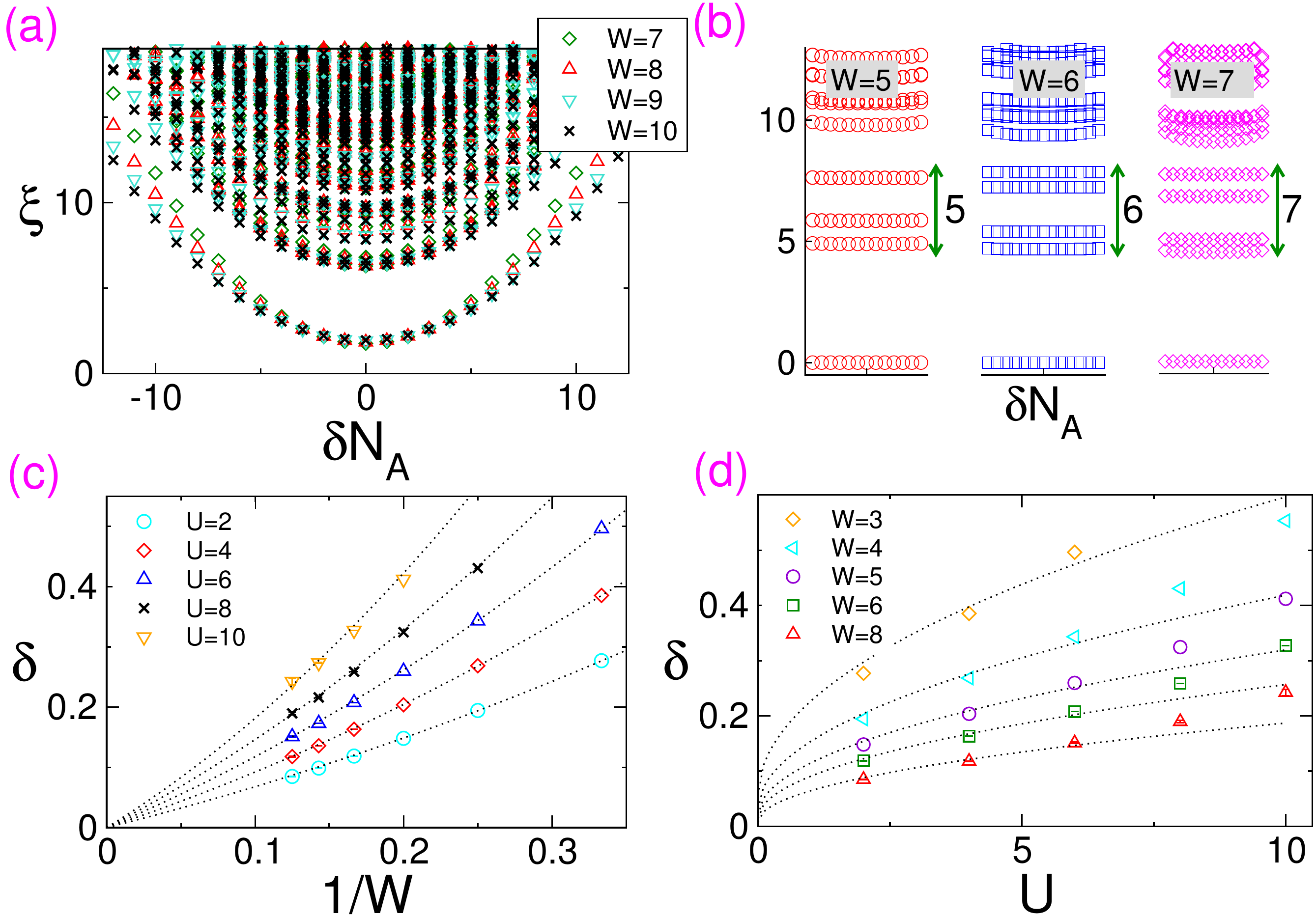}
\caption{ 
({\bf a}) ES in the superfluid phase   (DMRG data at $U=2$ for size $L/4=W$).  ({\bf b}) The same ES as in
({\bf a}) but after subtracting the contribution of the envelope. ({\bf c}) Spacing $\delta$ (defined in text and
Fig.~\ref{fig3}) plotted versus $1/W$; various $U$.  Dotted lines are fits to $A/W+B/W^2$.  ({\bf d}) $\delta$ as
function of $U$.  Dotted lines are fits to $\delta\sim \sqrt{U}$ (expected for $U{\to}0$).
}
\label{fig2}
\end{figure}
%%%%%%%%%%%%%%%%%%%%%%%%%%%%%%%%%%

\paragraph* {The ES in the superfluid phase ---}

In the superfluid phase ($U\lesssim16.739$~\cite{sansone-2008}) the ES looks dramatically different (Fig.~\ref{fig2}).  There is a clearer
separation between a low-lying ``envelope'' and the rest of the ES, but the envelope now has quadratic dependence on
$\delta{N_A}$, and there is only a single envelope level at each $\delta{N_A}$.  These features are due to the fact that
the underlying superfluid state has spontaneous breaking of $U(1)$ symmetry in the thermodynamic limit.
We can explain some of these features through a correspondence with the tower of states
spectrum \cite{lhuillier_arxiv2005}, which is the physical low-energy spectrum obtained when a system with spontaneously
broken continuous symmetry is placed in a finite volume.  Since the finite-size ES is plotted against quantum numbers
whose conservation is spontaneously broken only in the thermodynamic limit, it is naturally related to the TOS spectrum.
For the lower part of the ES, ${\mathcal H}_E \sim {\mathcal H}_{T}/T_E$, where ${\mathcal H}_{T}$ is the TOS Hamiltonian,
$T_E$ is an effective temperature given by $T_E=v_s/{\mathcal L}$ with $v_s$ the velocity of the gapless excitations
(for the Bose Hubbard this is the sound velocity), and ${\mathcal L}\approx W$ is the linear size of the system
\cite{met-grov-2011}.
The form of $T_E$ reflects the finite size behavior of the sound-wave gap.  For the Bose-Hubbard the TOS Hamiltonian is
${\mathcal H}_T\sim (\delta\hat{N})^2/(\chi V)$ where $\hat{N}$ is the total particle number operator and $\chi\equiv
dn/d\mu$ is the compressibility.  This suggests that ${\mathcal H_E} \sim \frac{W}{v_s\chi V_A}(\delta N_A)^2\sim (\delta
N_A)^2/(v_s\chi W)$.

\begin{figure}[t]
\includegraphics[width=1\columnwidth]{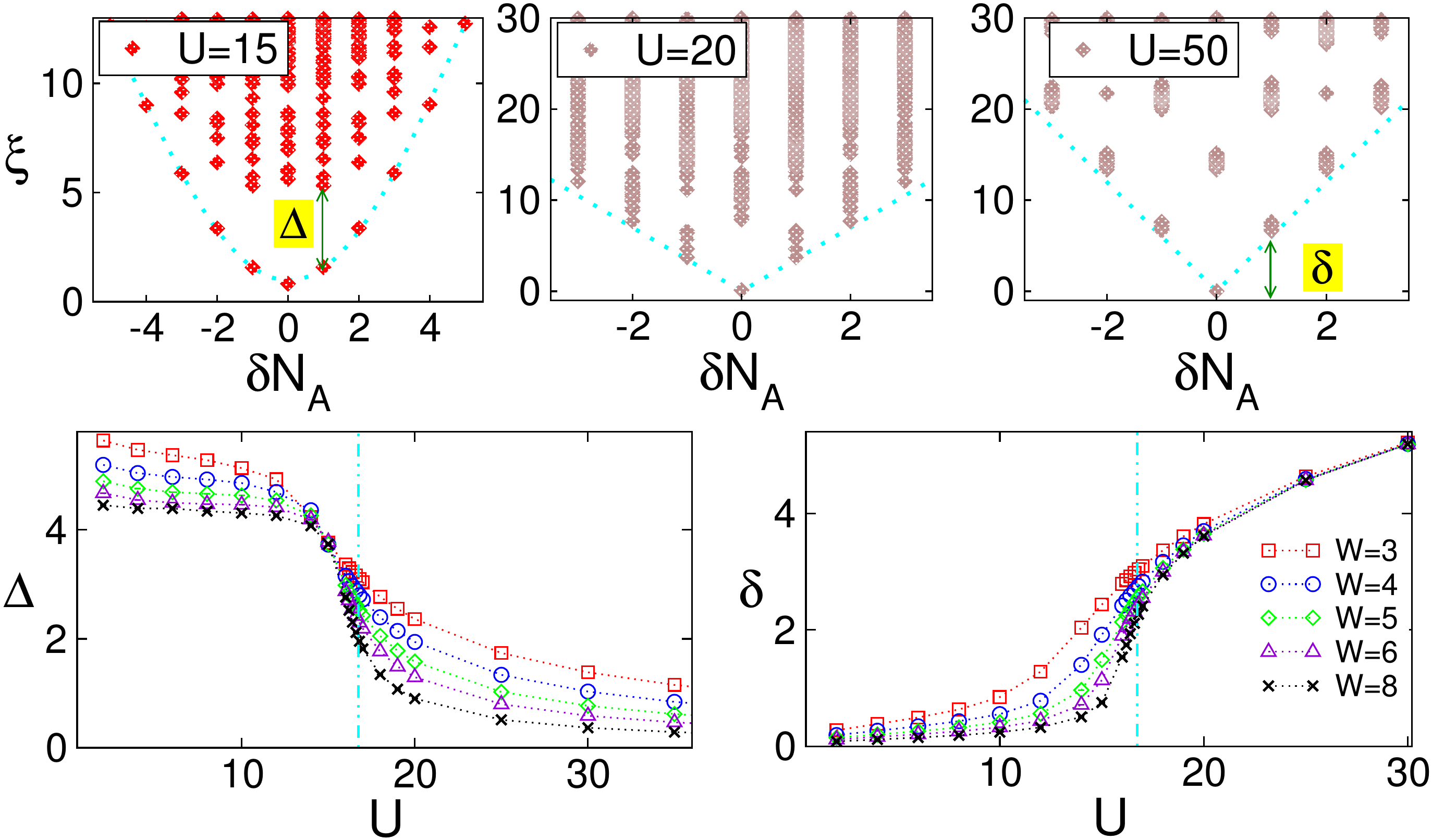}
\caption{ \textbf{(top)}  Restructuring of ES across  the Mott-superfluid transition  (DMRG data for $W=8=L/4$, equal
bipartitions).  The arrows explain the definitions of the quantities $\Delta$ and $\delta$.
\textbf{(bottom)}   $\Delta$ and  $\delta$  plotted versus $U$ for  several values of the boundary length $W$.  The vertical line
denotes the critical value $U_c$.
} 
\label{fig3}
\end{figure}

We characterize this scenario through the quantities $\Delta$ (``TOS gap'') and $\delta$ (``envelope
curvature''), defined pictorially in Fig.~\ref{fig3}.  Formally, $\delta\equiv\xi_{k=0}-\xi_0$ and
$\Delta\equiv \xi_{k=2\pi/W}-\xi_{k=0}$, with $\xi_{k=0}, \xi_{k=2\pi/W}$ in the lowest ES multiplet at $\delta N_A=1$,
while $\xi_0$ is the lowest ES level.  
In the thermodynamic limit ${\mathcal H}_E \sim {\mathcal O}(1/W)$, then $\delta\sim {\mathcal O}(1/W)$. Furthermore, in
the limit $U\to 0$ at fixed $W$, using $\chi^{-1}\sim U$ and $v_s\sim\sqrt{U}$ one obtains
$\delta\sim\sqrt{U}$. Finally, assuming that the lowest excitations above the ES envelope are sound-wave like one can
write ${\mathcal H}_E\sim [{\mathcal H}_T +{\mathcal H}_{sw}]/T_E$ where ${\mathcal H}_{sw}$ describes the sound wave
excitations.  Since $T_E\sim 1/W$ this suggests that the gap $\Delta$ remains finite in the thermodynamic limit,
although logarithmic decay cannot be ruled out \cite{met-grov-private}.

%%%%%%%%%%%%%%%%%%%%%%%%%%%%%%%%%%
%%%%%%%%%%%%%%%%%%%%%%%%%%%%%%%%%%

Numerical data in Figs.~\ref{fig2} and \ref{fig3} show a finite gap $\Delta$ for all $W$ and a quadratic behavior of
the ES envelope, supporting the tower of states picture.  Figs.~\ref{fig2} ({\bf c}) and \ref{fig2} ({\bf d}) show good
agreement with the predictions $\delta\sim1/W$ and $\delta\sim\sqrt{U}$.

In Fig.~\ref{fig2} ({\bf b}) and \ref{fig3} we notice also that the gap $\Delta$ hardly changes with $W$ (although a
$\sim{1}/\log{W}$ type of decay cannot be ruled out.)  Surprisingly, $\Delta$ is also constant as a function of $\delta
N_A$.  We also note that, above the gap, the ES levels possess further band-like structures with a band of $W$ levels of
width comparable to $\Delta$, slightly but distinctly separated from higher levels.  Field theory arguments involving
the dynamics of sound waves indicate similar structures \cite{met-grov-private}.

At the $U=0$ point, the ES can be calculated exactly \cite{supplement, ding-2009}.  There is a single ES level for
each $\delta{N_A}$, i.e., only the envelope survives and the rest of the ES is pushed to infinity.  The wavefunction for
$U=0$ is an exact Bose condensate; the highly symmetric situation is similar to the ferromagnetic case \cite{ferro}.

\paragraph*{The Mott-superfluid transition ---}

Across the phase transition ($U_c\approx 16.739$~\cite{sansone-2008}), $\Delta$ and $\delta$ show ``dual'' behaviors
(Fig.~\ref{fig3} bottom).  In the Mott insulating phase, $\Delta\sim 1/W^2$ (from \eqref{tight}
one has $W^2\Delta\sim\pi^2/6$), while in the superfluid $\Delta$ converges to a possibly nonzero value or vanishes
only logarithmically with $W$ \cite{met-grov-private}.  
In contrast, the $W$-dependence of $\delta$ is: $\delta\sim\mathrm{const.}$\ in the Mott insulator, $\delta\to 0$ in the
superfluid.  The observed $W$-dependence of $\delta$ suggests that this quantity might allow a scaling collapse,
similar to the 1D example of Ref.~\cite{DeChiara2012}. 
The closing of a quantity similar to $\delta$ at a quantum critical point has been reported in another 2D
example~\cite{Konik}.

%%%%%%%%%%%%%%%%%%%%%%%%%%%%%%%%%e
\begin{figure}[t]
\includegraphics*[width=0.99\linewidth]{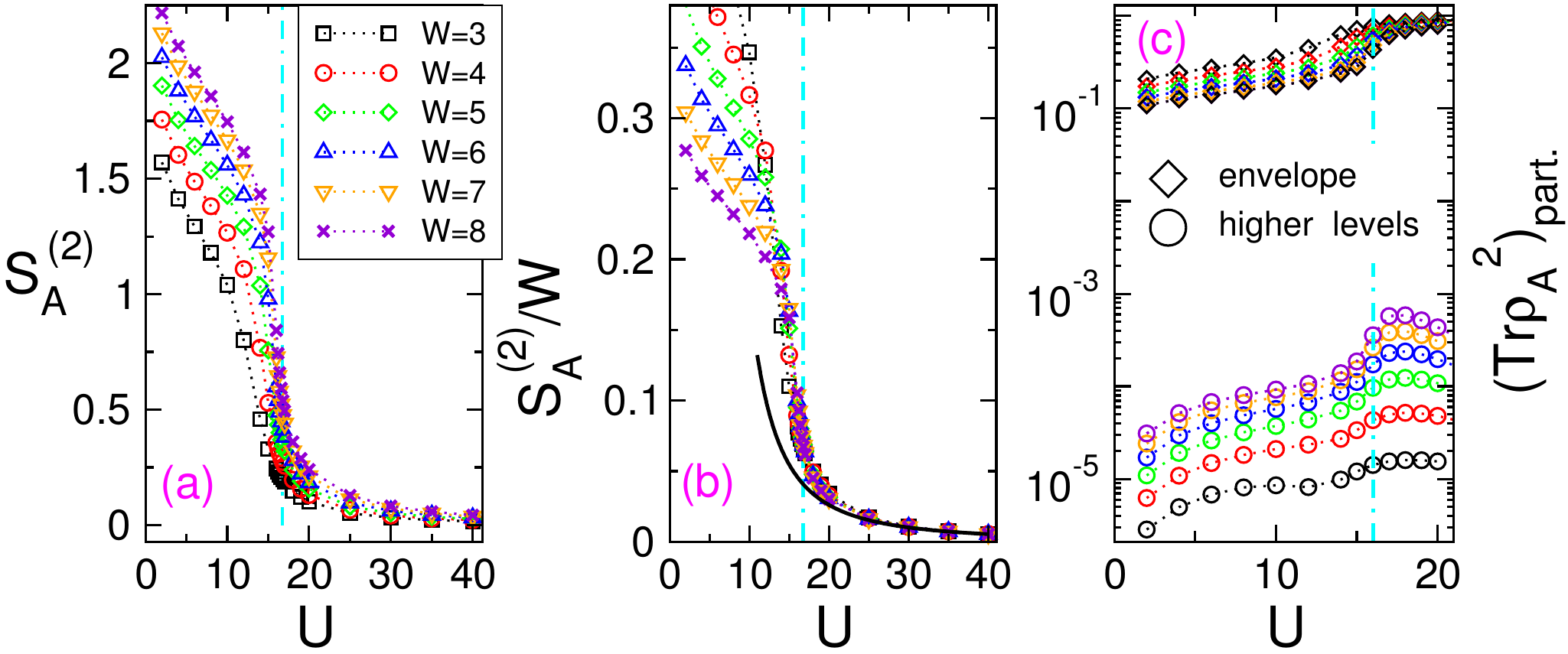}
\caption{ 
\textbf{(a,b)} Renyi entropy $S_A^{(2)}$ (DMRG data). In \textbf{(b)} the data of  \textbf{(a)} is rescaled by boundary
length $W$, and the continuous line is the perturbative result.
\textbf{(c)} Contributions to $\textrm{Tr}\rho_A^2$ from the envelope ES levels (rhombi) and from the higher levels
(circles).  Vertical lines in each panel marks the transition point.
}
\label{fig4}
\end{figure}
%%%%%%%%%%%%%%%%%%%%%%%%%%%%%%%%%%

\paragraph*{Entanglement entropies ---}

In the Mott phase, entropies can be constructed perturbatively in $1/U$, similar to other 2D gapped
systems \cite{kallin-2011, singh-2012, TagliacozzoEvenblyVidal_PRB09, vanLeeuwen_2DIsing_dmrg_PRB98}.
%
%% (See \cite{kallin-2011, singh-2012, TagliacozzoEvenblyVidal_PRB09, vanLeeuwen_2DIsing_dmrg_PRB98}
%% for similar considerations in other 2D gapped phases.)
%
%% Perturbative considerations for entanglement entropies in 2D gapped phases also appear in
%% Refs.\ \cite{kallin-2011, singh-2012, TagliacozzoEvenblyVidal_PRB09, vanLeeuwen_2DIsing_dmrg_PRB98}.
%
We first consider the single-copy entanglement, $\xi_0 = \lim\nolimits_{\gamma\to\infty}S_A^{(\gamma)}$ with
$S^{(\gamma)}\equiv-\log\textrm{Tr}\rho_A^\gamma/(\gamma-1)$ the Renyi entropy of index $\gamma$.  Up to
$\mathcal{O}(U^{-4})$ we find $\xi_0= cW$ with $c=4/U^2+488/U^4$, i.e.  the expected area law behavior.
Fig.\ \ref{fig1}(\textbf{i}) compares with DMRG data at $U=100$.  Remarkably, up to $\alpha=4$ the only dependence on
$W$ is the linear area law and higher powers in $W$ disappear in $\xi_0$ \cite{supplement}.
The same is true for the von Neumann entropy and the Renyi entropies with $\gamma=2,3$
\cite{supplement}.

In the superfluid ($0\le U<U_c$) the entropies are still expected to exhibit an area law, although, due to the symmetry
breaking, subleading logarithmic corrections arise \cite{kallin-2011, Song_LeHur_PRB2011, met-grov-2011}, i.e.
$S^{(\gamma)}_A\sim c_1W+c_2\log W$.  At $U=0$ there is pure logarithmic behavior $S_A\sim 1/2\log N_A$
\cite{supplement}.  Thus the area law coefficient vanishes as $U\rightarrow0$.  The entanglement entropies are dominated
by logarithmic corrections up to larger system sizes as $U$ decreases.

% The transition between these different regimes is illustrated in Fig.~\ref{fig4} for $S^{(2)}_A$. 
%
In Fig.~\ref{fig4} $({\bf a})$ we show $S_A^{(2)}$ for $3\le W\le 8$ as a function of $U$. A sharp change is visible at
the phase transition.  The data divided by $W$ (Fig.~\ref{fig4} $({\bf b})$) shows that in the Mott phase the entropy
follows an area law and the behavior is fully reproduced by the perturbative result within the expected regime.  In the
superfluid, $S_A^{(2)}$ shows strong finite size effects.  In Fig.~\ref{fig4} $({\bf c})$, we show the contribution to
$\textrm{Tr}\rho_A^2$ coming from the ES envelope (which is responsible for the subleading logarithmic
correction~\cite{met-grov-2011}) and from the ES levels above.
The contribution of the non-envelope levels is much lower, especially in the superfluid (about four orders of
magnitude).
%
%% This means that most of the contribution to $S_A^{(2)}$ comes from the envelope, which is responsible of the subleading
%% logarithmic correction~\cite{met-grov-2011}.

These results have implications for the efficiency of DMRG.  The logarithmic contribution is due to the replication of
the ES in every $\delta{N_A}$ sector, which would be suppressed if $\delta{N_A}$ was not a conserved quantity, as
happens in DMRG calculations where the $U(1)$ symmetry is enforced to be broken.  (See
Refs.~\cite{WhiteChernyshev_PRL07,Bauer2012} for analogous calculations in magnetic systems.) Our results thus provide a
more quantitative understanding of why explicitly symmetry-broken DMRG outperforms simulations exploiting the associated
conserved quantum numbers, when the underlying phase breaks a continuous symmetry.

\paragraph{Discussion ---}

We have presented a thorough analysis of the ES of the 2D Bose-Hubbard model, highlighting very different ES structures
in the Mott and superfluid phases.  In the Mott phase, the ES can be understood perturbatively from the $U\to\infty$
product state.  The ES is organized in a perturbative hierarchy and there are sub-structures reflecting the extended
nature of the boundary.  In the superfluid regime, the ES reflects the spontaneous symmetry breaking of the underlying
state, and is related to the tower of states known from finite-size studies of physical spectra.  We believe the major
features addressed in the two phases are generic for a wide range of gapped and gapless 2D systems.

Our results open up a number of research directions.  First, identifying the TOS in the ground state ES implies that one
can use ground state analysis to explore exotic symmetry breaking scenarios (e.g.~\cite{lhuillier_arxiv2005,
  exotic_sym_breaking_TOS}), avoiding the expensive requirement of calculating many eigenstates.  Second, our analysis
points to an advantage of DMRG calculations with enforced symmetry breaking, as explained above.  Exploration of such
techniques is yet to be attempted in detail.  Third, it would be interesting to explicitly see the TOS structure appear
for gapless phases where the symmetry breaking is more complex than the $U(1)$ symmetry broken in our case.  
%
%% The ES envelope would then be expected to have a nontrivial degeneracy structure, which would provide a more stringent
%% demonstration of the correspondence between the ES and the TOS spectrum.
%
Finally, the robust emergence of an area law in perturbation theory (cancellation of $W^{\alpha}$ terms with $\alpha>1$)
deserves to be better understood in a more general setting.

\acknowledgments

We acknowledge stimulating discussions with T. Grover, M. Metlitski, and R.R.P. Singh.
The DMRG simulations have been performed on machines of the platform "Scientific computing" at the University of
Innsbruck - supported by the BMWF.

\newpage

%%%%%%%%%%%%%%%%%%%%%%%%%%%%%%%%%%%%%%%%%%%%%%%%%%%%%%%
% definitions for Supplementary

\setcounter{figure}{0}   \renewcommand{\thefigure}{S\arabic{figure}}

\setcounter{equation}{0} \renewcommand{\theequation}{S.\arabic{equation}}

%%%%%%%%%%%%%%%%%%%%%%%%%%%%%%%%%%%%%%%%%%%%%%%%%%%%%%%

\begin{center}
\underline{\Large Supplementary Materials}
\end{center}

%%%%%%%%%%%%%%%%%%%%%%%%%%%%%%%%%%%%%%%%%%%%%%%%%%%%%%%%%%%%%%%
\section{Overview}

In these Supplements, we provide additional details and information on several aspects of the ES of the 2D Bose-Hubbard
model:

\begin{itemize}
\addtolength{\itemsep}{-2\itemsep}
%\addtolength{\labelwidth}{-\labelwidth}
%
\item We provide details on the perturbative expansions of the entanglement entropies in the Mott phase given in the
paper.  In particular we show that the expected area law behavior seems to be built into the structure of  perturbation theory.
\item  We provide perturbative expressions in the Mott phase, up to the first few orders, for the entanglement
Hamiltonian describing the levels of the ES envelope.  We demonstrate that up to the first few orders the entanglement
Hamiltonian has the form of a tight-binding $1D$ Hamiltonian for hard-core bosons, whose degrees of freedom are
localized at the boundary between the two subsystems (boundary locality).  By examining the sectors $\delta
N_A=1,2$, we show that while in the Mott phase the entanglement Hamiltonian is short range it becomes long range as the
Mott superfluid transition is approached.
\item  We discuss the  $U=0$ point.  We show that the ES contains only one level in each sector $\delta N_A$ for
which we provide analytic expressions.  The behaviors of the entanglement entropies and `gaps' ($\Delta$ and $\delta$)
are discussed and compared with the interacting case $U{\ne}0$.
\end{itemize}

%%%%%%%%%%%%%%%%%%%%%%%%%%%%%%%%%%%%%%%%%%%%%%%%%%%%%%%%%%%%%%%
\section{Entanglement entropies in Mott phase}

As discussed in the main text, the entanglement entropies can be calculated perturbatively at large $U$.  Here we
provide expressions up to fourth order in $1/U$.  This requires reduced density matrx eigenvalues $\lambda_i$ (or ES
levels $\xi_i$) up to order $\alpha=2$ as defined in Figure 1 (left panel) of the main text.

We find that all the ES levels depend only on the boundary length $W$ and not on the transverse size $L$, reflecting the
boundary local nature of the ES.  The perturbative expressions for ES levels contain terms nonlinear in $W$.
Remarkably, at the orders we have calculated, these cancel when one constructs the Renyi and von Neumann entropies,
giving rise to the expected area law ($\propto{W}$) behavior.

%%%%%%%%%%%%%%%%%%%%%%%%%%%%%%%%%%%%%%%%%%%%%%%%%%%%%%%%%%%%%%
\subsection{The reduced density matrix eigenvalues}

We denote the eigenvalues of $\rho_A$ at perturbative order $\alpha$ and particle number sector $\delta{N_A}$ by
$\lambda^{\{\alpha,\delta N_A\}}_i$, and the corresponding ES levels as $\xi_i^{\{\alpha,\delta N_A\}} =
-\ln\lambda^{\{\alpha,\delta N_A\}}_i$.  The index $i$ labels the levels within each group at order $\alpha$ and sector
$\delta{N_A}$.

%%%%%%%%%%%%%%%%%%%%%%%%%%%%%%%%%%%%%%%%%%%%%%%%%%%%%%%%%%%%%
\paragraph{\underline{$\alpha=0$}.}
%\paragraph{\underline{The dominant eigenvalue}.}

Up to fourth order in $1/U$ the dominant eigenvalue (corresponding to the lowest ES level $\xi_0$) is given by
\begin{equation}
\label{xi0}
\lambda^{\{0,0\}}=1-4W/U^2-(488W-8W^2)/U^4+{\mathcal O}
(U^{-5})
\end{equation}
As may be expected from the boundary locality of the ES, the eigenvalue depends only on the boundary length $W$, and not
on the cylinder length $L$ in the direction transverse to the boundary.  It is noteworthy that \eqref{xi0} contains
terms nonlinear in $W$.

%%%%%%%%%%%%%%%%%%%%%%%%%%%%%%%%%%%%%%%%%%%%%%%%%%%%%%%%%%
\paragraph{\underline{$\alpha=1$}.}
%\paragraph{\underline{First order eigenvalues}.}

There are $W$ levels at $\alpha=1$ in the sector $\delta N_A=\pm{1}$.  Up to fourth order in $1/U$ we find
\begin{multline}
\label{xi1}
\lambda^{\{1,\pm1\}}_j=\frac{2}{U^2}+\frac{148}{U^4}
-W\frac{8}{U^4}+\frac{48}{U^3}\cos k_j+\\\frac{224}{U^4}
\cos 2k_j+{\mathcal O}(U^{-5})
\end{multline}
with $k_j\equiv 2\pi{j}/W$ ($j=1,2\dots,W$). 

%% the momentum along the boundary direction which, due to the periodic boundary conditions, is a good quantum number
%% for $\rho_A$.

%%%%%%%%%%%%%%%%%%%%%%%%%%%%%%%%%%%%%%%%%%%%%%%%%%%%%%%%%%
\paragraph{\underline{$\alpha=2$}.}
%\paragraph{\underline{Second order eigenvalues}}

Finally, we need to know the expansion of $\lambda^{\{2,\pm 2\}}$ and $\lambda^{\{2,0\}}$.  At $\alpha=2$ order there
are $W(W-1)/2$ eigenvalues each in sectors $\delta N_A=\pm 2$ and $W(W-1)$ in sector $\delta N_A=0$  (Figure 1
of main article).  All these eigenvalues are degenerate up to fourth order: 
\begin{equation}
\label{xi2}
\lambda_i^{\{2,\pm 2\}}=\lambda_i^{\{2,0\}}=
\frac{4}{U^4}+{\mathcal O}(U^{-5})
\end{equation}

%%%%%%%%%%%%%%%%%%%%%%%%%%%%%%%%%%%%%%%%%%%%%%%%%%%%%%
\subsection{Entanglement entropies}

Using the perturbative results \eqref{xi0}, \eqref{xi1}, \eqref{xi2}, we derive the corresponding expressions for the 
entropies.

\paragraph{\underline{The single copy entanglement}.}

The single copy entanglement $S_A^{(\infty)}$ is obtained 
from~\eqref{xi0} as 

\begin{equation}
\label{single0}
S_A^{(\infty)}=-\log\lambda^{\{0,0\}}
\end{equation}
which has the expansion
\begin{multline}
\label{single}
S_A^{(\infty)}=W\frac{4}{U^2}+W\frac{488}{U^4}+W^2
\frac{976}{U^6}\\ -W^3\frac{4}{U^6}+W^3\frac{16}{3U^6} +
{\mathcal O}(U^{-6})
\end{multline}
Since \eqref{xi0} is correct only up to terms ${\mathcal O}(U^{-5})$,  then in~\eqref{single} only the terms up to fourth
order are meaningful.  It is remarkable that up to fourth order~\eqref{single} contains only linear terms in $W$ (area
law), even though in~\eqref{xi0} a quadratic term $\sim W^2$ is present. This term cancels when expanding the logarithm
in~\eqref{single0}.  The $\sim{U}^{-6}$ terms apparently violate the area law but are not meaningful at this order and
have to be discarded. 
%
%% Such contributions violationg the area law arise from the behavior of the multiplicities (as $W^\alpha$) (see
%% Fig.~\ref{fig1} {\bf left panel}).
%
At present we are unaware of any proof that higher powers of $W$ must vanish at every perturbative order.

%Notice that 
%the multiplicity of the eigenvalues grows exponentially with the
%perturbative order $\alpha$ (as $W^\alpha$). This would give 
%non linear terms (violating the area law) proportional to $W^\alpha$.

\paragraph{\underline{Renyi entropies}.}

The expansion for the other Renyi entropies $S_A^{(\gamma)}$ ($\gamma\ge 2$) is obtained using
\begin{equation}
\label{renyi}
S_A^{(\gamma)}=-\frac{1}{n-1}\log\textrm{Tr}\rho_A^{\gamma} \, .
\end{equation}
For $\gamma=2$ we have
\begin{equation}
\textrm{Tr}\rho_A^2=1-W\frac{8}{U^2}-W\frac{968}{U^4}+W^2\frac{32}{U^4}
+{\mathcal O}(U^{-5})  \, . 
\end{equation}
Note again the presence of quadratic terms which arise from the multiplicity ($\sim W^2$) of $\lambda_i^{(2,\pm 2)}$ and
$\lambda_i^{(2,0)}$. Plugging in~\eqref{renyi} and expanding the logarithm we get
\begin{equation}
\label{S_2}
S_A^{(2)}=W\frac{8}{U^2}+W\frac{968}{U^4}+{\mathcal O}(U^{-5}) \, .
\end{equation}
Again, we find the area law behavior up to meaningful order, due to the cancellation of $W^2$ terms as observed above
for the single copy entanglement.  Similar observations hold for the $\gamma=3$ Renyi entropy: 
\begin{equation}
\textrm{Tr}\rho_A^3=1-W\frac{12}{U^2}-W\frac{2928}{U^4}+W^2\frac{72}{U^4}
+{\mathcal O}(U^{-5})  \, .
\end{equation}

\paragraph{\underline{Von Neumann entropy}.}

Up to order four the von Neumann entropy is
\begin{multline}
S_A\equiv -\textrm{Tr} \rho_A\log\rho_A
\\
= -\lambda^{\{0,0\}}\log\lambda^{\{0,0\}}  
-\sum\limits_{\delta\in\{-1,1\}}\sum\limits_{i=1}
^W\lambda^{\{1,\delta\}}_i\log\lambda^{\{1,\delta\}}_i
\\
-\sum\limits_{\delta\in\{-2,2\}}\sum\limits_{i=1}
^{W(W-1)/2}\lambda^{\{2,\delta\}}_i\log\lambda^{\{2,\delta\}}_i\\
-\sum\limits_{i=1}
^{W(W-1)}\lambda^{\{2,0\}}_i\log\lambda^{\{2,0\}}_i
\end{multline}
Using \eqref{xi0}, \eqref{xi1}, \eqref{xi2}, and the identity 
\[
\sum_{q=1}^W\cos(2\pi/W rq)=W\delta_{r,0\mod W} \, ,
\] 
with $r$ an integer, we obtain
\begin{multline}
S_A ~=~ 4W\Big[\frac{1}{U^2}-\frac{1}{U^2}\log \frac{2}{U^2}
-\frac{16}{U^4}\log\frac{2}{U^2}-\frac{146}{U^4}\Big]
\\ +~\ord(U^{-5})  \, .
\end{multline}
As observed for the Renyi entropies, although~\eqref{xi0}, 
\eqref{xi1}, \eqref{xi2} contain non linear terms in $W$, 
the von Neumann entropy exhibits the area law behavior.

%%%%%%%%%%%%%%%%%%Entanglement H. for the envelope%%%%%%%%
\section{Entanglement Hamiltonian for the ES envelope}

On the ES envelope in the Mott phase (Figure 1 of main article), at each nonzero $\delta{N_A}$ there is a cluster of states
which are described by an entanglement Hamiltonian and show corresponding dispersions.  Below we give details for these
entanglement Hamiltonians.

\paragraph{\underline{The $\delta{N_A}=1$ sector}.}

\begin{figure}[t]
\includegraphics[width=.95\columnwidth]{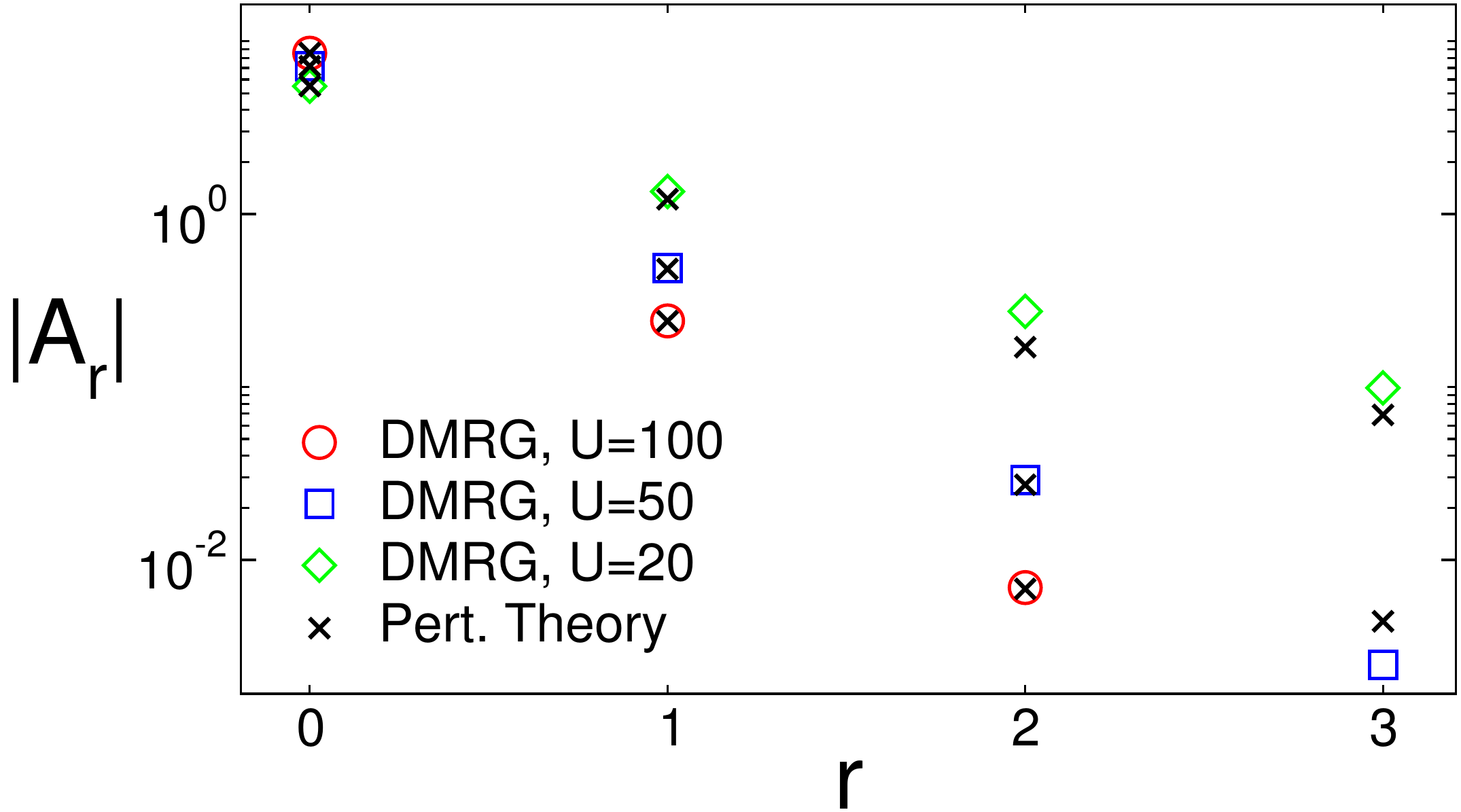}
\caption{ \label{fig6}
Entanglement Hamiltonian for the envelope ES multiplet in the sector $\delta{N_A}=1$. We plot the absolute value of the
hopping amplitudes $|A_r|$ versus the hopping range $r$, on log-linear scale.  The empty symbols are DMRG data in the
Mott insulating phase (for a system of size $W=L/2=8$).  Perturbation theory results are shown with crosses.
}
\end{figure}

The entanglement Hamiltonian is given by the single particle hopping Hamiltonian
\begin{equation}
\label{single_EH}
{\mathcal H}_E=\sum_{ir} A_r(a^\dagger_i
a_{i+r}+h.c.)
\end{equation}
where $i=1,2,\dots,W$ labels the sites near the boundary between the two subsystems and $r=0,1,2\dots$ is the hopping
range.  Up to ${\mathcal O}(U^{-4})$ the hopping amplitudes are found to be
\begin{align}
\label{EH_d1}
& A_0\equiv\log(U^2/2)+2(2W+10)/U^2+{\mathcal O}(U^{-4})\\
\nonumber & A_1\equiv-12\Big[1/U+(2W-25)/(2U^3)\Big]+
{\mathcal O}(U^{-4})\\
\nonumber & A_2\equiv-34/U^2+{\mathcal O}(U^{-4})\\
\nonumber& A_3\equiv226/U^3+{\mathcal O}(U^{-4})
\end{align}
In Fig.~\ref{fig6} we compare the theoretical predictions \eqref{EH_d1} with DMRG data. We consider a system of
dimensions $W=L/2=8$ and $U=20,50,100$.  In Fig.\ \ref{fig6} we show $|A_r|$ (for $r=0,1,2,3$) obtained
from~\eqref{EH_d1}.  We also plot the hopping amplitudes $|A_r|$ obtained by fitting the DMRG data to the single
particle tight binding dispersion
\begin{equation}
\xi_m=A_0+2\sum\limits_{j=1}^3A_j\cos\frac{2\pi}{W}m j   \; .
\end{equation} 
The perturbative expressions \ref{EH_d1} describe the DMRG data extremely well for $U=50,100$.  Not surprisingly, the
agreement is less perfect for $U=20$ (the transition to the superfluid regime is at $U\approx 17$).  The hopping
amplitudes  $|A_r|$ are seen to decay exponentially in the Mott phase with the hopping range $r$, but longer range
hopping becomes more and more relevant as the superfluid transition is approached.

%%%%%%%%%%%%%%%%%%%%%%%%%%%%%%%%%%
%%%%%%%%%%%%%%%%%%%%%%%%%%%%%%%%%%

\paragraph{\underline{The $\delta N_A=2$ sector.}}

We provide the first two non zero orders (in $1/U$) for the entanglement Hamiltonian describing the levels of the ES
envelope at $\delta N_A=2$. The block in the reduced density matrix corresponding to this sector can be written as
\begin{equation}
\label{red_2}
\rho_A(\delta N_A=2)=\frac{4}{U^4}\sum\limits_{i=1}^{W}
\left[\frac{1}{2}a_i^\dagger a_i+\frac{12}{U}
(a_i^\dagger a_{i+1}+h.c.)\right]+\dots
\end{equation}
where $i$ labels the sites of subsystem $A$ near the boundary, and $a_i^\dagger$ are hard-core boson creation
operators.  The dots denote higher order corrections.  The corresponding block in the entanglement Hamiltonian
${\mathcal H}_E\equiv-\log\rho_A$ is
\begin{equation}
\label{EH_2}
{\mathcal H}_E(\delta N_A=2)=4\log\frac{U}{\sqrt{2}}+{\mathcal H}'
+\dots
\end{equation}
with
\begin{equation}
\label{H_hcb}
{\mathcal H}'\equiv-\frac{12}{U}
\sum\limits_{i=1}^{W}(a_i^\dagger a_{i+1}+h.c.)  \, .
\end{equation}
Apart from the constant shift $4\log(U/\sqrt{2})$ the entanglement Hamiltonian is tight-binding Hamiltonian for two
hard-core bosons, \eqref{H_hcb}.  Thus the ES levels in the envelope multiplet at $\delta N_A=2$ can be obtained from
the energy spectrum of two free hard-core bosons on a periodic chain.  The single particle spectrum of~\eqref{H_hcb} is
given by
\begin{displaymath}
\epsilon_l=\left\{
\begin{array}{cccc}
-\frac{24}{U}\cos\frac{2\pi}{W}(l+\frac{1}{2}) & \textrm{if}& W 
&\textrm{even}\\
\\
-\frac{24}{U}\cos\frac{2\pi}{W} l & \textrm{if} & W & \textrm{odd}\\
\end{array}
\right.
\end{displaymath}
with $l=0,1\dots,W-1$.  The ES, i.e., the spectrum of \eqref{EH_2}, is obtained as
\begin{equation}
\label{ES_2}
4\log(U/\sqrt{2})+\epsilon_{l_1}+\epsilon_{l_2}
\end{equation}
with $l_1<l_2=0,1,\dots,W-1$.

%%%%%%%%%%%%%%%%%%%%%%%%%%%%%%%%%%
\begin{figure}[t]
\includegraphics[width=.95\columnwidth]{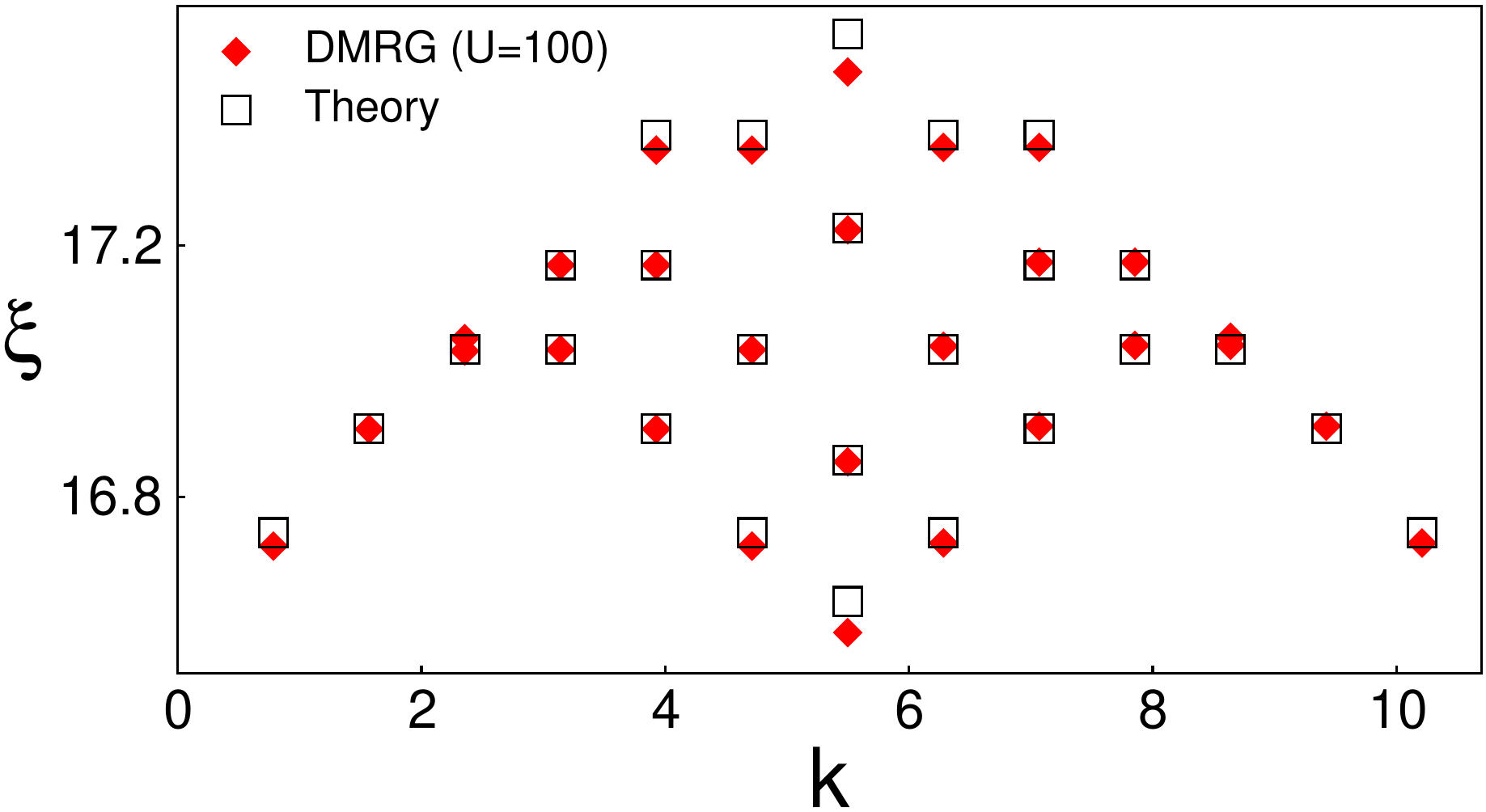}
\caption{ \label{fig5}
Entanglement dispersion in the envelope ES levels at $\delta N_A=2$.  ES levels $\xi$ are plotted against the momentum
$k$ along the boundary direction.  Filled symbols are DMRG data at $U=100$ for $W=L/2=8$.  Empty symbols denote the
theoretical prediction~\eqref{ES_2}.
}
\end{figure}
%%%%%%%%%%%%%%%%%%%%%%%%%%%%%%%%%%

In Fig.~\ref{fig5} we show the ES levels in the envelope multiplet at $\delta N_A=2$ (DMRG data for $U=100$ and
$W=L/2=8$). We also show the theoretical result~\eqref{ES_2} plotting the ES versus the momentum along the boundary
$k=2\pi l_1/W+2\pi l_2/W$.  Since in our DMRG implementation we did not have access to the momentum, we assigned the
momentum labels such that the data matches the theoretical prediction.  (The procedure is not unique because ES levels
with different momenta can be degenerate, but this is not important for present purposes.)  Apart from small deviations
due to higher order corrections in $1/U$, DMRG data are in agreement with~\eqref{ES_2}.

\paragraph{\underline{The general case $\delta N_A>2$}.}

The result~\eqref{ES_2} can be extended to all the other clusters of levels in the ES envelope.  The block in the
entanglement Hamiltonian describing the envelope states at any $\delta N_A$ is given (up to the first subleading order)
by the the tight-binding Hamiltonian for $|\delta N_A|$ hard-core bosons:
\begin{equation}
\label{ESenv_EH}
{\mathcal H}_E(\delta N_A)=2|\delta N_A|\log\frac{U}{\sqrt{2}}
+{\mathcal H}'
+\dots
\end{equation}
with ${\mathcal H}'$ given again by~\eqref{H_hcb}.

%%%%%%%%%%%%%%%ES OF THE BOSE CONDENSATE%%%%%%%%%%%%%%%%%%%%%
\section{ES at the $U=0$ point}
%\section{ES in the limit $U\to 0$}

Given a system of $N$ non interacting bosons ($U=0$), the ground state is an exact Bose condensate.  All the bosons are in the
single particle state
\begin{equation}
\label{sin_par_st}
|\psi\rangle=\sum_{i=1}^{V}c_i b^\dagger_i|0\rangle
\end{equation}
with $V=LW$ the total number of sites of the system.  (Since we consider unit
filling we also have $N=V$.) Note that the ``usual'' Bose condensate in the
state at zero momentum ($k=0$) corresponds to the choice
$c_i=\frac{1}{\sqrt{V}} \quad \forall i$ in~\eqref{sin_par_st}.  This would
give the correct ground state for non-interacting bosons on a torus.  Since we
are working on a cylindrical geometry this is not exactly correct in our case.

The ES obtained from the Bose condensate in the state~\eqref{sin_par_st} contains only one level in each sector with
fixed $N_A$ that is given by
\begin{equation}
\label{U_0_ES}
\xi^{(N_A)}=-\log\left[\big(\sum_{i\in A}|c_i|^2\big)^{N_A}
\big(\sum_{i\in B}|c_i|^2\big)^{N_B}
\frac{N!}{N_A!N_B!}\right].
\end{equation}
The ES~\eqref{U_0_ES} is continuously connected to the ES envelope upon switching on the interaction
(at $U\ne 0$).  However, the gap $\delta$ has very different behavior at $U=0$ and for $U\ne0$, as
we will see below.  

In other words, in the limit $U\to 0$ only the ES envelope survives and the higher parts of the ES are pushed to
infinity.  This implies that the gap $\Delta$ diverges ($\Delta\to\infty$) in the $U\to0$ limit.  In our DMRG data
(Figure 3 bottom left in main article), we do not see a divergence for $U$ as low as 2.  This could be because the
divergence starts at smaller $U$ at these sizes, or that DMRG simulations with finite boson number cutoff per site is
not able to capture this divergence.

For the toric geometry (when $c_i=1/\sqrt{V}$ ) the ES~\eqref{U_0_ES} becomes
\begin{equation}
\label{U_0_ES_tor}
\xi^{(N_A)}=-\log\left[\frac{V_A^{N_A}V_B^{N_B}}{V^{N}}\right]
+\log\left[\frac{N_A!N_B!}{N!}\right]  \, .
\end{equation}
For our cylindrical geometry one obtains~\eqref{U_0_ES_tor} only if block $A$ is half of the system.  Since this is the
bipartition that we considered in the paper (and we expect that the physics remains unchanged for different
bipartitions) we can use ~\eqref{U_0_ES_tor}. Note that in the limit $\delta{N_A} \equiv N_A-N/2\ll N$ the ES is exactly
parabolic
\begin{eqnarray}
\label{U_0_ES_per}
\xi^{(N_A)}=-\frac{1}{2} \log\left(\frac{2}{V\pi}\right)+
\frac{2}{V}(\delta N_A)^2  \, .
\end{eqnarray}

Strikingly, \eqref{U_0_ES_per} implies $\delta \sim 1/V$, whereas at $U>0$ the presence of sound waves (and of the ES
levels above the envelope) implies $\delta\sim 1/W\sim1/\sqrt{V}$ (as discussed in the main text).  

As a final remark, we mention that the entanglement entropy for non-interacting bosons does not obey the area law and in
the limit $N\to\infty$ is given by
\begin{equation}
S_A\approx \frac{1}{2}\log N_A \, .
\end{equation} 
This is consistent with the physics that the envelope in the gapless phase gives a logarithmic contribution to the
entanglement entropies, while the area law is due to sound waves.  Since sound waves are missing at $U=0$, the
coefficient of the area law term also vanishes in this limit.


\begin{thebibliography}{99}

\bibitem{AmicoFazioOsterlohVedral_RMP08} L. Amico, R. Fazio,
  A. Osterloh, and V. Vedral, Rev.\ Mod.\ Phys.\ {\bf 80}, 517 (2008).



\bibitem{topological}
H.~Li and F.~D.~M.~Haldane,  Phys.\ Rev.\ Lett.\ {\bf 101}, 010504 (2008).  
%
N.~Regnault, B.~A.~Bernevig, F.~D.~M.~ Haldane, Phys.\ Rev.\ Lett. {\bf 103}, 016801 (2009).
%
N.~Bray-Ali, L.~Ding, and S.~Haas, Phys.\ Rev.\ B {\bf 80}, 180504(R) (2009). 
%
L.~Fidkowski, Phys.\ Rev.\ Lett.\ {\bf 104}, 130502 (2010).
%
A.~M.~L\"auchli, E.~J.~Bergholtz, J.~Suorsa, and M.~Haque, Phys.\ Rev.\ Lett.\ {\bf 104}, 156404 (2010).
%
R.~Thomale, A.~Sterdyniak, N.~Regnault, and B.~A.~Bernevig, Phys.\ Rev.\ Lett.\ {\bf 104}, 180502 (2010).
%
H.~Yao and X.~L.~Qi, Phys.\ Rev.\ Lett.\ {\bf 105}, 080501 (2010).
%
E.~Prodan, T.~L.~Hughes, and B.~A.~Bernevig, Phys.\ Rev.\ Lett.\ {\bf 105}, 115501 (2010).
%
F.~Pollmann, A.~M.~Turner, E.~Berg, M.~Oshikawa, Phys.\ Rev.\ B, {\bf 81}, 064439 (2010).
%
M.~Kargarian and G.~A.~Fiete, Phys.\ Rev.\ B, {\bf 82}, 085106 (2010).
%
A.~M.~Turner, Y.~Zhang, A.~Vishwanath, Phys.\ Rev.\ B, {\bf 82}, 241102R (2010).
%
Z.~Papic, B.~A.~Bernevig, and N.~Regnault, Phys.\ Rev.\ Lett.\ {\bf 106}, 056801 (2011).
%
L.~Fidkowski, T.~S.~Jackson and I.~Klich, Phys.\ Rev.\ Lett.\ {\bf 107}, 036601 (2011).
%
J.~Dubail, and N.~Read, Phys.\ Rev.\ Lett.\ {\bf 107}, 157001 (2011). 
%
J.~Schliemann, Phys.\ Rev.\ B {\bf 83}, 115322 (2011). 
%
T.~L.~Hughes, E.~Prodan, B.~A.~Bernevig, Phys.\ Rev.\ B, {\bf 83}, 245132 (2011).
%
N.~Regnault and B.~A.~Bernevig, Phys.\ Rev.\ X {\bf 1}, 021014 (2011).
%
X.~L.~Qi, H.~Katsura, and A.~W.~W.~Ludwig, Phys.\ Rev.\ Lett.\ {\bf 108}, 196402 (2012).
%
D. Poilblanc, N. Schuch, D. Perez-Garcia, and J.I. Cirac, Phys. Rev. B {\bf 86}, 014404 (2012);
%
N. Schuch, D. Poilblanc, J.~I.~Cirac, and D.~Perez-Garcia, arXiv:1210.5601 (unpublished).

\bibitem{fisher-89} M.~P.~A.~Fisher, P.~B.~Weichman, G.~Grinstein, and D.~S.~Fisher,
Phys.\ Rev.\ B {\bf 40}, 546 (1989).

\bibitem{jaksch-98} D.~Jaksch, C.~Bruder, J.~I.~Cirac, C.~W.~Gardiner, P. Zoller, 
Phys.\ Rev.\ Lett.\ {\bf 81} 3108 (1998).

\bibitem{alba-2011} V.~Alba, M.~Haque and A.~M.~L\"auchli, Phys.\ Rev.\ Lett.\ {\bf 108}, 227201 (2012).

\bibitem{met-grov-2011} M.~A.~Metlitski and T.~Grover, arXiv:1112.5166.

\bibitem{lhuillier_arxiv2005}
C.~Lhuillier, arXiv:cond-mat/0502464v1 (unpublished). 

\bibitem{poil-2011} 
J.~I.~Cirac, D.~Poilblanc, N.~Schuch, and F.~Verstraete, Phys.\ Rev.\ B {\bf
  83}, 245134 (2011).  
I. Peschel and M.-C. Chung, EPL {\bf 96}, 50006 (2011).

\bibitem{supplement} Additional detail is provided in the Supplementary Materials.


\bibitem{sansone-2008} B.~Capogrosso-Sansone, S.~G.~S\"oyler, N.~Prokof'ev, and B.~Svistunov, Phys.\ Rev. \ A {\bf 77}, 015602 (2008).


\bibitem{met-grov-private} M.~A.~Metlitski and T.~Grover, private communication. 


\bibitem{ding-2009} W.~Ding and K.~Yang, Phys.\ Rev.\ A {\bf 80}, 012329 (2009).

\bibitem{ferro} V.~Alba, M.~Haque, and A.~M.~L\"auchli;\, J.~Stat.\ Mech.\ P08011 (2012).  
%
V.~Popkov and M.~Salerno, Phys.\ Rev\ A {\bf 71}, 012301 (2005).
%
M.~Salerno and V.~Popkov, Phys.\ Rev\ E {\bf 82}, 011142 (2010).
%
O.~A.~Castro-Alvaredo and B.~Doyon, J.\ Stat.\ Mech.\ P02001 (2011).


\bibitem{DeChiara2012}
G. De Chiara, L. Lepori, M. Lewenstein, and A. Sanpera, Phys. Rev. Lett. {\bf 109}, 237208 (2012).

\bibitem{Konik} A.J.A.~James and R.M. Konik, arXiv:1208.4033 (unpublished).

\bibitem{vanLeeuwen_2DIsing_dmrg_PRB98}
M.~S.~L. du Croo de Jongh and J.~M.~J.\ van Leeuwen, Phys.\ Rev.\ B {\bf 57}, 8494 (1998).

\bibitem{TagliacozzoEvenblyVidal_PRB09} 
L.~Tagliacozzo, G.~Evenbly, and G.~Vidal, Phys.\ Rev.\ B {\bf 80}, 235127 (2009).

\bibitem{singh-2012} R.~R.~P.~Singh, R.~G.~Melko, J.~Oitmaa, Phys. Rev. B {\bf 86}, 075106 (2012).

\bibitem{kallin-2011} 
A.~B.~Kallin, M.~B.~Hastings, R.~G.~Melko, and R.~R.~P.~Singh, Phys.\ Rev.\ B {\bf 84}, 165134 (2011).

\bibitem{Song_LeHur_PRB2011} 
H.~F.~Song, N.~Laflorencie, S.~Rachel, and K.~Le~Hur, Phys.\ Rev\ B {\bf 83}, 224410 (2011).

\bibitem{WhiteChernyshev_PRL07} S.~R.~White and A.~L.~Chernyshev, Phys.\ Rev.\ Lett.\ {\bf 99}, 127004 (2007).

\bibitem{Bauer2012} B. Bauer, P. Corboz, A.M. L\"auchli, L. Messio, K. Penc, M. Troyer, and F Mila, Phys. Rev. B {\bf
  85}, 125116 (2012).

\bibitem{exotic_sym_breaking_TOS}  B.~Bernu, C.~Lhuillier, and L.~Pierre, 
Phys. Rev. Lett. {\bf 69}, 2590 (1992).
%
A.~L\"auchli, J.~C.~Domenge, C.~Lhuillier, P.~Sindzingre, and M.~Troyer,
Phys.\ Rev.\ Lett.\ {\bf 95}, 137206 (2005).
%
N.~Shannon, T.~Momoi, and P.~Sindzingre, Phys.\ Rev.\ Lett.\ {\bf 96}, 027213 (2006).
%
K.~Penc, and A.~M.~L\"auchli, Springer Series in Solid-State Sciences {\bf 164}, 331 (2011).
\end{thebibliography}
\end{document}